\documentclass[12pt]{iopart}
\usepackage{iopams}

\newcommand{\bra}{\left\langle}
\newcommand{\ket}{\right\rangle}

\newcommand{\Ps}{P_{\rm s}}
\newcommand{\Ple}{P_{\rm le}}

\newcommand{\ep}{\epsilon}

\begin{document}

\title[Large deviation functionals in fluctuating hydrodynamics
]{A perturbation theory for large deviation 
functionals in fluctuating hydrodynamics}

\author{Shin-ichi Sasa\footnote[3]
{To whom correspondence should be addressed
(sasa@jiro.c.u-tokyo.ac.jp).} }

\address{Department of Pure and Applied Sciences, University of Tokyo, 
Komaba, Tokyo 153-8902, Japan}

\begin{abstract}
We study a large deviation functional of density fluctuation
by analyzing  stochastic non-linear diffusion equations driven 
by the difference between the densities fixed at the boundaries. 
By using a fundamental equality that yields the fluctuation 
theorem, we first relate the large deviation functional
with a minimization problem. We then develop a  perturbation 
method for solving the problem. In particular, by performing an
expansion  with respect to the average current, we derive 
the lowest order expression for the deviation from the local 
equilibrium part. This expression implies that the 
deviation is written as the space-time integration 
of the excess entropy production rate during the most probable 
process of generating the fluctuation that corresponds to 
the argument of the large deviation functional. 
\end{abstract}




\section{Introduction}

The nature of macroscopic fluctuations in equilibrium systems 
is determined by thermodynamics. Mathematically, this implies
that the large deviation functional for the fluctuations 
is given by a thermodynamic function. This relation is called 
Einstein's formula, and it is an important consequence of 
equilibrium statistical mechanics. 

Apart from thermodynamic systems, there are many systems in which 
macroscopic fluctuations are studied. In general, it appears sound
to consider that properties of macroscopic fluctuations for non-equilibrium
systems depend on the systems under consideration. However, 
from an optimistic viewpoint, it may be expected  that there 
exists a class of non-equilibrium systems in which the large deviation 
functional has a physical correspondence with  macroscopic deterministic 
laws such as those in thermodynamics and hydrodynamics. 

Recently, researches speculating the existence of such a system
have been presented. In one approach, an extended form of 
thermodynamics is explored before considering fluctuations 
\cite{OonoPaniconi98}. If it is operationally and consistently constructed, 
this extended function 
might provide a large deviation functional. Although there are some 
successful examples \cite{HayashiSasa03,SST}, it turned out  a 
severe restriction is required to extend a thermodynamic framework
in macroscopic systems \cite{SST}.

Another approach utilized  a new variation principle
called the {\it additivity principle} \cite{DerridaLebowitzSpeer01,
DerridaLebowitzSpeer02b,BodineauDerrida04}. Indeed, with the aid 
of this principle,  non-trivial forms of  large deviation 
functionals  were exactly derived for some models. Although 
the discovery of this principle stimulates us to expect it
to possess generality, there is no evidence that it is useful
for a wide class of non-equilibrium systems. 

Considering the above-mentioned  difficulties in  
constructing a general framework, we have decided that 
concrete expressions of large 
deviation functionals should be investigated for several models. 
With this motivation, we attempt to develop a 
perturbation method for calculating large deviation functionals.
In particular, we analyze a fluctuating hydrodynamic model that 
describes the  density fluctuations in systems that are 
attached to particle reservoirs at the boundaries (see Sec. \ref{pre}). 
Even for this simple model, a perturbation method has never been 
proposed, except for the approximation method by which large 
deviation functionals
are treated as a quadratic form \cite{Spohn}.  This approximate
expression can be obtained by the analysis of a linearized 
equation, but it is not appropriate for our purpose because 
we wish to know the physical correspondence of large 
deviation functionals. 

In this paper, 
we expand a large deviation functional with respect 
to the average current, maintaining the nature of
large fluctuations. Here, it should be noted that 
the large deviation functional can be exactly derived for equilibrium 
cases. This exact derivation is due to  a detailed 
balance property. Thus, in order to develop a 
perturbation method with respect to the average current, 
we need to find a useful relation that corresponds to an extended 
version of the detailed balance property. Fortunately, 
a fundamental relation that recovers the detailed balance 
condition in equilibrium cases is known to be related to  
the fluctuation theorem 
\cite{ECM,GC,Kurchan,Maes,Crooks}.  Using this relation, in Sec. \ref{for},
we formulate a minimization problem to which a perturbation method can 
be applied for determining the large deviation functional.
In Sec. \ref{per}, we calculate the large deviation functional 
of the lowest order with respect to the average current. 
Section \ref{remark} is devoted to concluding remarks. 

\section{Preliminaries}\label{pre}

\subsection{Model}

We consider a one-dimensional system in contact with particle 
reservoirs at the boundaries $x=0$ and $x=L$. Let $\rho(x,t)$ be 
the coarse-grained density field. We assume that $\rho(x,t)$ 
obeys the stochastic partial differential equation
\begin{equation}
\partial_t \rho=\partial_x(D(\rho)\partial_x \rho+\xi)
\label{evol}
\end{equation}
with  Gaussian white noise  $\xi$ that satisfies 
\begin{equation}
\bra \xi(x,t) \xi(y,s) \ket=2  \sigma(\rho(x)) T \delta(x-y)\delta(t-s),
\label{noise0}
\end{equation}
where $D$ in (\ref{evol}) and $\sigma$ in (\ref{noise}) represent
the diffusion constant and the  conductivity in the system
under consideration, respectively.  $T$ is the temperature, and
the Boltzmann constant is set to unity. It is straightforward
to extend our analysis for $d(\ge 2)$-dimensional systems,
but for simplicity we focus on the one-dimensional case.

Since we are interested in the large deviation functional of
the density field, we set $\epsilon=1/L$ and introduce a new space-time
coordinate $(x', t')=(\epsilon x, \epsilon^2 t)$.
With this coordinate, we define a new field 
$\rho'( x',  t')=\rho(x, t)$ and noise 
$\xi'( x',  t')=\epsilon^{-1} \xi(x, t)$. Then, 
$ \rho'( x',  t')$ satisfies the same equation 
as (\ref{evol}) with prime symbols. Omitting
the prime symbols, we obtain (\ref{evol}) with 
\begin{equation}
\bra \xi(x,t) \xi(y,s) \ket=2 \epsilon
\sigma(\rho(x)) T\delta(x-y)\delta(t-s).
\label{noise}
\end{equation}
We impose the boundary conditions $\rho(0,t)=\rho_0$ and 
$\rho(1,t)=\rho_1$, where we assume that $\rho_0 \le \rho_1$ 
without loss of generality. 

\subsection{Notation}

Since a few types of functional dependence appear in this paper,
we use the following notations in order to avoid confusion. 
Formally, $\rho$ represents the density field as a function 
of $(x,t)$, where $ 0 \le x \le 1$ and $0 \le t  \le \infty$.
However, as in the usual practice in physics, we write $\rho(x,t)$ 
as $\rho$ if the argument can be easily guessed. On the contrary, 
when a quantity $g$ depends on the function $\rho$, we 
explicitly write the functional dependence as $g(\rho())$. 
Thus, for example, $D(\rho)$ in (\ref{evol}) does not represent 
the functional dependence of $\rho$, but it should be read as 
$D(\rho(x,t))$. Furthermore, the density field as a function of $x$
is denoted by $\hat \rho$. In a similar manner, 
$h(\hat \rho)$ represents $h(\hat \rho(x))$ and $g(\hat \rho())$
represents the functional dependence of $\hat \rho$. We also
use the expression $\rho(\ ,0)=\hat \rho()$ as an identity
for functions of $x$. That is, this means that 
$\rho(x,0)=\hat \rho(x)$ for any value of $x$. 

\subsection{Question}

We study  statistical properties in the steady state 
of this model. The average current $\bar J$ and 
average density profile $\bar \rho(x)$ are obtained 
from the relation 
\begin{equation}
D(\bar \rho)\partial_x \bar \rho = -\bar J
\label{average}
\end{equation} 
with the boundary conditions $\bar\rho(0)
=\rho_0$ and $\bar \rho(1)=\rho_1$. However, it has been known 
that fluctuations around the average value exhibit  non-trivial 
behaviors such as non-local correlations \cite{Spohn}. 
Here, let $\Ps(\hat \rho())$
be the stationary distribution of density fluctuations. In 
the limit $\epsilon \to 0$ ($L \to \infty$ in the original problem),
it is characterized by the leading order expression
\begin{equation}
\Ps(\hat \rho())\simeq\exp\left[-\frac{1}{\epsilon} I(\hat \rho()))
\right],
\end{equation}
where $I(\hat \rho())$ is called the large deviation functional. 

Despite the simplicity of the model, it is quite difficult to calculate 
the large deviation functional for general non-equilibrium  cases. To date,
only for the special case where $ D=1$ and $\sigma(\rho)=\rho(1-\rho)$, 
it was derived as a variational form by using the Hamilton-Jacobi formulation
for the model described by (\ref{evol}) with (\ref{noise}) 
\cite{Bertinietal02}. 
It should be noted that this large 
deviation functional is identical to the exact solution for a simple 
exclusion process in contact with particle reservoirs at 
the boundaries \cite{DerridaLebowitzSpeer01}.
However, we find that the special technique for solving the Hamilton-Jacobi 
equation cannot be used for other functional forms of 
$D(\rho)$ and $\sigma(\rho)$.

Here, let us consider the problem from a physical viewpoint. 
We first review the result for the equilibrium case $\rho_0=\rho_1$.
According to Einstein's formula, the large deviation 
functional is written as
\begin{equation}
I (\hat \rho()) =\beta [F(\hat \rho()) -F( \bar \rho()) ]
\label{I-F}
\end{equation}
with
\begin{equation}
F(\hat \rho())=\int_0^1 dx[f(\hat \rho(x))- \mu \hat \rho(x)],
\label{Fdef0}
\end{equation}
where $\beta$ is the inverse temperature; $f(\rho)$, the 
free energy density; $\mu$, the chemical potential
defined by $\mu=f'(\rho_0)$. Hereafter, the prime represents
the derivative with respect to the density.
Furthermore, with regard to the intensity of density 
fluctuations 
$\chi(= \bra \left[\int_0^1 dx (\rho(x,t)-\rho_0)\right]^2 \ket/\epsilon)$, 
we know that the equality  $ f'' =T \chi^{-1}$ and the Einstein 
relation $D \chi =\sigma T$ hold (see the review in Ref. \cite{HS-PA}). 
These lead to 
\begin{equation}
f''(\rho) = \frac{D(\rho)}{\sigma(\rho)}.
\label{hdef}
\end{equation}
In this manner, we have determined the large deviation 
functional without any calculation.  On the contrary, 
without any physical consideration, we can derive the 
large deviation functional in the form of (\ref{I-F}) 
using (\ref{Fdef0}) and (\ref{hdef}). 
(See the argument below (\ref{Ndef}) in Sec. \ref{statio}.)

The physical argument developed above cannot
be applied to non-equilibrium cases ($\rho_0 < \rho_1)$
because thermodynamics for non-equilibrium
steady states has not yet been established. 
Nevertheless, 
it is expected that the large deviation functional 
for non-equilibrium cases might have a correspondence 
with the local equilibrium 
form, where the function $F()$ given in (\ref{Fdef0}) 
is replaced with
\begin{equation}
F_{\rm le}(\hat \rho())
=\int_0^1 dx [f(\hat \rho(x)) -\mu(x) \hat \rho(x)].
\label{local}
\end{equation}
The chemical potential in this expression depends on $x$,
and its functional form is given by the local equilibrium 
thermodynamics. That is, 
\begin{equation}
\mu(x)=f'(\bar \rho(x)).
\label{cpot}
\end{equation}
Further, considering the relation
\begin{equation}
D(\rho)\partial_x \rho=\sigma(\rho) \partial_x f'(\rho)
\label{D-sigma}
\end{equation}
derived from (\ref{hdef}), we obtain
\begin{equation}
\bar J =-\sigma(\bar \rho) \partial_x \mu.
\end{equation}
This implies that $\sigma$ is indeed the conductivity.

Here, it should be noted that the large deviation function is
{\it not} equal to $\beta[F_{\rm le}(\hat \rho())-F_{\rm le}(\bar \rho())]$,
but there is a deviation from the local equilibrium part. 
We then express the function $F$ in (\ref{I-F}) as 
\begin{equation}
F(\hat \rho())=F_{\rm le}(\hat \rho())+N(\hat \rho() ).
\label{Nint}
\end{equation}
At present, the specific objective of this study is to find a
physical interpretation of $N (\hat \rho())$ by obtaining 
its expression.  

\section{Formal analysis}\label{for}

\subsection{Fundamental equality}

We first define 
\begin{equation}
F_0(\hat \rho() )=\int_0^1  dx f(\hat \rho(x)),
\end{equation}
where $f$ is given in (\ref{hdef}). 
We then rewrite (\ref{evol}) as the continuity equation
\begin{equation}
\partial_t \rho+\partial_x j=0,
\label{cont}
\end{equation}
with 
\begin{equation}
j=-\sigma(\rho) \partial_x \frac{\delta F_0}{\delta \rho(x)}+\xi,
\end{equation}
where we have used (\ref{D-sigma}).
Since $\xi$ obeys the Gaussian process, the probability distribution 
of $\rho(\ ,t)$, $0 <t \le \tau$, with fixed $\rho(\ ,0)$, is
written as 
\begin{equation}
{\cal P}(\rho()) \simeq \exp
\left[-\frac{\beta}{4\epsilon}
\int_0^\tau  dt\int_0^1 dx \frac{1}{\sigma(\rho)}\left(
j+\sigma(\rho) \partial_x\frac{\delta F_0}{\delta \rho(x)}\right)^2
\right],
\label{pathprob}
\end{equation}
where the current $j$ is connected with $\rho$ by the 
continuity equation given in (\ref{cont}). Note that the so-called 
Jacobian term does not appear in the leading order 
in the limit $\epsilon \to 0$. 

We next assume that the configuration $\hat \rho$ at $t=0$ obeys
the local equilibrium distribution 
\begin{equation}
\Ple ( \hat \rho ())
\simeq \exp\left[ -\frac{\beta}{\epsilon}
 \left( F_{\rm le}( \hat \rho () ) -  F_{\rm le}( \bar \rho () )
\right) \right].
\label{p0}
\end{equation}
Let $A( \rho())$ be an arbitrary function of $\rho()$,
and let $\bra A \ket$ represent the average of $A$ 
with respect to the path probability distribution 
$\Ple (\hat \rho ){\cal P}( \rho() )$.

In the equilibrium case $(\rho_0=\rho_1)$, the model satisfies
the detailed balance condition. This property is also known as stochastic 
reversibility, and it can be formalized by introducing the time-reversed 
trajectory of $\rho$ that is denoted by $\rho^{\dagger} $. This can
be explicitly written as $ \rho^\dagger(x, t)=\rho(x,\tau-t)$. 
In terms of $\rho^\dagger$, 
the stochastic reversibility is expressed as a symmetry property 
$\bra A \ket = \bra A^\dagger  \ket$, where  
$ A^\dagger (\rho() ) \equiv A(\rho^\dagger())$. 
The equality $\bra A \ket = \bra A^\dagger  \ket$ is not valid for 
non-equilibrium systems. We then attempt to derive its extended relation.
The key equality for deriving this relation is as follows:
\begin{eqnarray}
\frac{ {\cal P}(\rho() )}{ {\cal P}(\rho^\dagger () )} 
&=& \exp
\left( -\frac{\beta}{\epsilon}
\left[ F_0(\rho(\ , \tau))-F_0(\rho(\ , 0) ) \right. \right.\nonumber \\
& & + \left. \left. 
\int_0^\tau dt (\mu(1) j(1,t)-\mu(0)j(0,t)) 
    \right]
    \right),
\label{LDB}
\end{eqnarray}
which is obtained by direct calculation. 
Using this, we calculate 
\begin{eqnarray}
\bra A \ket &=& \int {\cal D}\rho
\Ple ( \rho(\ , 0) ){\cal P}( \rho()) A( \rho() )  \nonumber \\
           &=& \int {\cal D}\rho
\Ple ( \rho(\ , 0)) {\cal P}( \rho()) 
\frac{ \Ple ( \rho^\dagger(\ , 0)) }{\Ple ( \rho(\ , 0))  }
\frac{ {\cal P}(\rho^\dagger() )}{ {\cal P}(\rho () )} 
A^\dagger( \rho() )  \nonumber \\
&=& 
\bra  A^\dagger \exp
\left( \frac{\beta}{\epsilon}
\int_0^\tau  dt \int_0^1 dx  (\partial_x \mu) j 
\right)\ket.
\label{FT}
\end{eqnarray}
Since the relation $\bra A \ket =\bra A^\dagger \ket$ is 
derived in the equilibrium case ($\partial_x \mu=0$), 
(\ref{FT}) is regarded as an extended
form of the stochastic reversibility. It has been known that this
equality yields many non-trivial relations, including the Green-Kubo 
relation, Kawasaki's non-linear response relation, and 
the fluctuation theorem \cite{Crooks}. 

\subsection{Stationary distribution}\label{statio}

Following the method presented in Ref. \cite{Crooks}, 
we derive an expression for  the stationary distribution 
$P_{\rm s}(\hat \rho())$ based on the relation in (\ref{FT}).
The probability distribution of the configuration $\hat \rho $ 
is given by 
\begin{equation}
P_{\rm s}(\hat \rho())=\lim_{\tau \to \infty}
\bra \delta( \rho(\ , \tau)-  \hat \rho()) \ket.
\end{equation}
Substituting $A(\rho()) = \delta( \rho(\ , \tau) - \hat \rho() )$
in (\ref{FT}), we obtain
\begin{equation}
P_{\rm s}( \hat \rho() ) = \Ple( \hat \rho() )
\int_{\rho(\ ,0)= \hat \rho() } 
{\cal D}\rho \e^{-\frac{\beta}{4\epsilon}
\Sigma(\rho())},
\label{ps}
\end{equation}
with 
\begin{equation}
\Sigma(\rho())=
\int_0^\infty  dt \int_0^1 dx 
\left[ \frac{1}{\sigma(\rho)}
(j+D(\rho)\partial_x \rho)^2 
- 4j (\partial_x \mu) 
\right].
\label{Idef}
\end{equation}
Substituting (\ref{I-F}), (\ref{Nint}), and (\ref{p0}) into 
(\ref{ps}), and applying the limit $ \epsilon \to 0 $, we
obtain 
\begin{eqnarray}
N(\hat \rho()) &=& 
\frac{\beta}{4} \min_{ \rho(); \rho(\ , 0)=\hat \rho() }
\Sigma( \rho()  )  .
\label{Ndef} 
\end{eqnarray}
In equilibrium cases, using the condition 
$\partial_x \mu=0$, we can  trivially solve the minimization
problem as  $N(\hat \rho())=0$ 
because $\Sigma( \rho()) \ge 0$ for any $\rho()$ and 
the solution of the diffusion equation with 
$\rho(\ , 0)=\hat \rho()$ yields $\Sigma = 0$. 
On the contrary, the minimization problem is not easily
solved in non-equilibrium cases, as shown below. 
Note that the expression of the large deviation functional
in terms of the minimization over trajectories was presented
in Ref. \cite{Bertinietal02, Eyink, Eyink2}.  
However, to the best of our knowledge, 
the expression of $N(\hat \rho () )$ 
given in (\ref{Ndef}) has never been reported. 

In order to determine the trajectory minimizing $\Sigma$
on the right-hand side of (\ref{Ndef}),
we consider the variation $\rho(x,t) \to \rho(x,t)+\delta \rho(x,t)$.
We first assume that the trajectory  $\rho()$ that minimizes $\Sigma(\rho())$ 
satisfies 
\begin{equation}
\lim_{t \to \infty}  \rho(x ,t)=\bar \rho(x).
\label{t-inf:rho}
\end{equation}
Further, defining a new variable
\begin{equation}
\partial_x u=(j+D(\rho) \partial_x \rho)/\sigma(\rho), 
\label{udef}
\end{equation}
we calculate $\delta \Sigma=\Sigma(\rho()+\delta \rho())-\Sigma(\rho())$ as 
\begin{equation}
\delta \Sigma=\int_0^\infty  dt \int_0^1 dx 
\{  [ -\sigma'(\rho) (\partial_x u)^2-2D(\rho) (\partial_x^2 u)](\delta \rho)
   +2 (\partial_x u) (\delta j)  \}.
\label{var}
\end{equation}
From (\ref{cont}), $\delta j$ is related to $\delta \rho$ in the
form 
\begin{equation}
\delta j(x,t)=-\partial_t \partial_x \int dy \Delta^{-1}(x,y)\delta\rho(y,t),
\label{delj}
\end{equation}
where $\Delta^{-1}(x,y)$  is the Green function satisfying 
$\partial_x^2 \Delta^{-1}(x,y)=\delta(x-y)$, and 
$\Delta^{-1}(0,y)=\Delta^{-1}(1,y)=0$.
The substitution of (\ref{delj}) into the variational equation $\delta
\Sigma=0$ yields 
\begin{equation}
2 \int_0^1 dy (\partial_y \Delta^{-1}(y,x) )\partial_t \partial_y u(y,t)
=\sigma'(\rho) (\partial_x u)^2+2D(\rho) \partial_x^2u.
\label{uevol0}
\end{equation}
To this point, $u(0,t)$ and $u(1,t)$ can be any function 
of time. We here assume that the trajectory minimizing $\Sigma$
is obtained in a class of trajectories that satisfy
\begin{equation}
u(0,t)=u_0 \quad {\rm and} \quad u(1,t)=u_1,
\label{bc:u}
\end{equation}
where $u_0$ and $u_1$ are constants in time. 
Then, (\ref{uevol0}) becomes 
\begin{equation}
\partial_t u =-\sigma'(\rho) (\partial_x u)^2/2-D(\rho) \partial_x^2u.
\label{uevol}
\end{equation}
Furthermore, (\ref{cont}) and (\ref{udef}) lead to
\begin{equation}
\partial_t \rho =  \partial_x(D(\rho)\partial_x \rho
-\sigma(\rho) \partial_x u).
\label{revol}
\end{equation}
In this manner, with the two assumptions (\ref{t-inf:rho}) and (\ref{bc:u}),
we have found that 
the trajectory minimizing $\Sigma$ satisfies a set of equations
(\ref{uevol}) and (\ref{revol}). Here, it should be noted 
that the difference  
between  (\ref{pathprob}) and (\ref{Idef})
is the term $j \partial_x \mu$ in (\ref{Idef}) and that the 
variational equation for (\ref{Idef}) does not depend on 
this term. Thus, (\ref{uevol}) and (\ref{revol}) also 
determine the most probable trajectory under given boundary 
conditions  in the limit $\ep \to 0$. Indeed, (\ref{uevol}) 
and (\ref{revol}) are identical to the Hamiltonian equation
providing the most probable trajectories \cite{Bertinietal02}. 

Using this property, we consider the minimization problem.
First, let $\rho^{\rm F}$ be the solution of the deterministic 
equation $\partial_t \rho =\partial_x (D(\rho) \partial_x \rho)$
with $\rho(x,0)=\hat \rho(x)$. Then, $(\rho,u)=(\rho^{\rm F},{\rm const.})$
is one solution of the variational 
equation given in (\ref{uevol}) and (\ref{revol}).
This solution describes the most probable relaxation
behavior,  provided that $\rho(\ , 0)=\hat \rho()$,
but it yields $\Sigma \to \infty$ unless $\partial_x \mu=0$. 
(See (\ref{Idef}).)  Therefore, this most probable trajectory 
is not relevant in the minimization problem. 

In order to  seek  another solution, we further assume 
that the trajectory minimizing $\Sigma$ satisfies 
\begin{equation}
\lim_{t \to \infty} u(x,t) = u_*(x).
\label{t-inf:u}
\end{equation}
Then, $j$ approaches $j_*$ in this limit, where 
\begin{equation}
j_*=-D(\bar \rho) \partial_x \bar \rho+\sigma(\bar \rho)\partial_x u_*.
\label{31}
\end{equation}
Substituting this expression into  (\ref{uevol}), we obtain 
$j_*=\pm \bar J$.  When $j_*=\bar J$, we derive $\partial_x u_*=0$,
which corresponds to the case $\rho=\rho^{\rm F}$. We thus
pay attention to the solution for the case $j_*=- \bar J$. Then,
(\ref{31}) leads to 
\begin{equation}
\partial_x u_*(x)=2 \partial_x \mu,
\label{32}
\end{equation}
and we can easily confirm that $\Sigma  < \infty$ 
for this solution. 

The physical interpretation of this solution is obtained 
by considering the equality  $j_*=-\bar J$. 
Let $\rho^{\rm B}(x,t)=\rho(x,-t)$ for the solution. 
Then, the current for $\rho^{\rm B}$ is $\bar J$  at $t = -\infty$.
Further, defining 
\begin{equation}
u^{\rm B}(x,t)=-u(x,-t)+2f'(\rho(x,-t)),
\end{equation}
we can directly confirm that $(\rho^{\rm B}, u^{\rm B})$ 
satisfies (\ref{uevol}) and (\ref{revol}) with
the condition $(\rho^{\rm B}, u^{\rm B}) \to (\bar \rho(x), 0)$
as $t \to -\infty$.  
Thus, $\rho^{\rm B}(x,t)$ describes the most probable 
process of generating the fluctuation $\hat \rho()$ at $t=0$
starting  from $\bar \rho(x)$ at $t=-\infty$. Note that 
the boundary values $u^{\rm B}(0,t)=u^{\rm B}(1,t)=0$
are consistent with the condition imposed in Ref. \cite{Bertinietal02}. 


\section{Perturbation theory}\label{per}

\subsection{Idea}

Before developing a perturbation theory, 
we consider a numerical method by which 
we can calculate $N(\hat \rho() )$. 
First of all, we must solve (\ref{uevol}) and (\ref{revol}) 
under the conditions $(\rho,u) \to (\bar \rho,u_*)$ as 
$t \to \infty$ and $\rho(\ ,0)=\hat \rho( )$. Note that $u(x,0)$ 
should be determined  such that $(\rho,u) \to (\bar \rho,u_*)$ as 
$t \to \infty$. However, it is quite difficult 
to perform a shooting method for partial differential 
equations. 

Instead of such a direct method, we 
devise the following recursive procedure.
Initially, we set 
\begin{equation}
u^{(0)}(x,t)=u_*(x).
\end{equation}
Then, for $n \ge0$, we numerically solve 
\begin{equation}
\partial_t \rho^{(n)} =   
\partial_x( D(\rho^{(n)})\partial_x \rho^{(n)}-\sigma(\rho^{(n)})
\partial_x u^{(n)} )
\label{rrevol} 
\end{equation}
with $\rho^{(n)}(x,0)=\rho(x)$, and next we consider 
\begin{equation}
\partial_t u^{(n+1)} =-\sigma'(\rho^{(n)}) 
(\partial_x u^{(n+1)})^2/2-D (\rho^{(n)})\partial_x^2u^{(n+1)},
\label{urevol} 
\end{equation}
with $u^{(n+1)}(x,\infty)=u_*(x)$. Here, solving (\ref{urevol}),
we set $\tilde u(x,-t)=u(x,t)$. Then, $\tilde u$ satisfies
\begin{equation}
\partial_t \tilde u^{(n+1)} =\sigma'(\rho^{(n)}) 
(\partial_x \tilde u^{(n+1)})^2/2+D (\rho^{(n)})\partial_x^2 \tilde u^{(n+1)},
\label{u-revol2} 
\end{equation}
for $-\infty \le t \le 0$. Because $\tilde u^{(n)}(x,-\infty)=u_*(x)$, 
we can have a standard algorithm of the  forward time evolution 
starting from $u_*(x)$, 
and as the result $\tilde u^{(n)}(x,0)=u^{(n)}(x,0)$ is determined.
If we confirm the convergence $\rho^{(n)}(x,t) \to  \rho(x,t) $ and 
$u^{(n)}(x,t) \to  u(x,t)  $ within the desired numerical accuracy,
we obtain an approximate solution for (\ref{uevol}) and (\ref{revol})
under the conditions $(\rho,u) \to (\bar \rho,u_*)$ as 
$t \to \infty$ and $\rho(\ ,0)=\hat \rho( )$.

\subsection{Perturbative expansion}

The method we considered above is also useful for developing 
a perturbation theory. As the simplest example, 
we derive $N( \hat \rho()) $ up to the order of $\bar J^2$.  
Checking the $\bar J$ dependence 
of $N$ given in (\ref{Ndef}), we find that the $O(\bar J^2)$ contribution 
can be ignored in the estimation of $u$ and $j$. Considering this,
we investigate the recursion equation. First, we have 
$ u^{(0)}(x,t)= u_*(x)$ and 
\begin{equation}
\partial_t \rho^{(0)} =    
\partial_x(D(\rho^{(0)})\partial_x \rho^{(0)}-\sigma(\rho^{(0)}) 
\partial_x u_* )
\label{r-revol-hat} 
\end{equation}
with $\rho^{(0)}(x,0)=\hat \rho(x)$. Next, we solve
\begin{equation}
\partial_t u^{(1)} = -D (\rho^{(0)})\partial_x^2 u^{(1)}
-\sigma'(\rho^{(0)}) (\partial_x u^{(1)})^2/2
\label{u-revo-hatl} 
\end{equation}
under the condition $\lim_{t \to \infty} u^{(1)}(x,\infty)=u_*(x)$.
Substituting $u^{(1)}(x,t)=u^{(0)}(x,t)+v^{(1)}(x,t)$
into this equation and expanding in terms of $\bar J$, we obtain
\begin{equation}
v^{(1)}(x,t)=O(\bar J^2).
\end{equation}
Since $u^{(1)}(x,t)=u_*+O(\bar J^2)$,
we have arrived at the fixed point solution by ignoring 
the $O(\bar J^2)$ contribution in $u$ and $\rho$. 
Substituting this solution into (\ref{Idef}),  we obtain the
expression of $N(\rho())$ as 
\begin{equation}
N( \hat \rho() )=\beta \int_0^\infty  dt \int_0^1 dx (\partial_x \mu)
[D(\rho^{(0)})\partial_x\rho^{(0)} -\sigma(\rho^{(0)})\partial_x\mu]
+O(\bar J^3),
\label{Nexp}
\end{equation}
with (\ref{r-revol-hat}) and $\rho^{(0)}(x,0)=\hat \rho(x)$.

Now, let us recall the physical interpretation of the
solution. As discussed in the last paragraph of Sec. \ref{for},  
$\rho^{\rm B}(x,t)=\rho^{(0)}(x,-t)$ describes the most probable 
process for generating the fluctuation $\hat \rho()$ 
at $t=0$ starting from $\bar \rho(x)$ at $t=-\infty$. 
Further, from (\ref{r-revol-hat}), it is found that $J^{B} =D(\rho^{B})
\partial_x \rho^{\rm B}-\sigma(\rho^{\rm B}) \partial_x u_* $ 
represents the particle current in this most probable process. 
Using this current, we rewrite $N(\hat \rho())$ as
\begin{equation}
N( \hat \rho())=-\beta \int_{-\infty}^0 dt \int_0^1 dx 
(\partial_x \mu) [-J^{\rm B}-\sigma(\rho^{\rm B} )\partial_x\mu]+O(\bar J^3).
\label{Nexp2}
\end{equation}
Here, $-\beta (\partial_x \mu) J^{\rm B}$ is the entropy production rate 
observed in the process, while 
$ \beta \sigma(\rho^{\rm B} ) (\partial_x \mu)^2 $ 
is interpreted as the entropy production rate under the assumption
that the steady state with the density  $\rho^{\rm B}(x,t)$
is locally realized in a space-time point $(x,t)$.  We call the latter 
{\it quasi-steady entropy production}.  We then interpret 
$\beta (\partial_x \mu) [-J^{\rm B}-\sigma(\rho^{\rm B} )\partial_x\mu]$
as an excess entropy production ratio. With this interpretation,
we can regard $N( \hat \rho())$ as the space-time integration of 
the excess entropy absorption rate during the most probable process of
generating the fluctuation $\hat \rho()$ at $t=0$ starting from $\bar \rho(x)$ 
at $t=-\infty$. (Note the minus sign before the space-time 
integration.)

\subsection{Small fluctuations}

When we focus on small fluctuations, we can make another approximation by 
which a quantitative calculation becomes possible. 
Indeed, substituting $\rho(x,t)=\bar \rho(x)+\phi(x,t)$
into (\ref{Nexp}), we obtain 
\begin{equation}
N(\hat \rho())=-\frac{\bar J^2 A(\rho_0)}{2\sigma(\bar \rho_0)^2}
\int_0^\infty  dt \int_0^1 dx (\phi(x,t))^2
+O(\bar J^3, \phi^3),
\label{Nexp1-2}
\end{equation}
with
\begin{equation}
A(\rho_0)=\sigma''(\rho_0)-\sigma'(\rho_0)D'(\rho_0)/D(\rho_0).
\end{equation}
Furthermore, since we ignore the terms of $O(\bar J^3 )$ in $N$, 
we can assume that $\phi$ satisfies 
$\partial_t  \phi=\partial_x (D(\rho_0) \partial_x  \phi)$
with the initial condition $\phi(x,0)=\hat \rho(x,t)-\bar\rho$ 
and the boundary conditions $\phi(0,t)=\phi(1,t)=0$.
Here, considering that $\rho_1-\rho_0 \simeq O(\bar J)$, we find that 
$\rho_0$ in (\ref{Nexp1-2}) and the boundary conditions can 
be replaced with an arbitrary  value in $[\rho_0,\rho_1]$.
This approximation leads to
\begin{equation}
\Ps( \hat \rho()) \simeq \exp
\left[-\frac{\beta}{2\epsilon} 
\int_0^1 dy \int_0^1 dy' L(y,y') \phi(y)\phi(y')
\right].
\end{equation}
Here, we have defined 
\begin{equation}
L(y,y') =\frac{D(\bar \rho(y))}{\sigma(\bar \rho(y))}\delta(y-y')
-\frac{\bar J^2 A(\rho_0)}{\sigma(\rho_0)^2}
\int_0^\infty dt \int_0^1 dx G(x,y,t)G(x,y',t) ,
\label{Ldef}
\end{equation}
where $G(x,y,t)$ is the Green function that satisfies
\begin{equation}
[\partial_t -D(\rho_0) \partial_x^2]G(x,y,t)=0
\end{equation}
when $t >0$, $G(x,y,0)=\delta(x-y)$, and 
$G(0,y,t)=G(1,y,t)=0$. An expression that is essentially the same
as that in (\ref{Ldef}) was derived by
the analysis of the linearized equation around
the steady solution $\bar \rho$ \cite{Spohn}. 

Furthermore, note that the space-time integration
appearing in (\ref{Ldef}) is calculated as 
\begin{equation}
\int_0^\infty dt \int_0^1 dx G(x,y,t)G(x,y',t) 
=
\frac{y'(1-y)}{2D(\rho_0)}\theta(y-y')+(y \leftrightarrow y'),
\end{equation}
where $\theta(\ )$ is Heaviside's step function. 
Then, from the Gaussian nature of the fluctuations,
we derive 
\begin{eqnarray}
\bra \phi(y)\phi(y') \ket
&=&  
\epsilon T \frac{\sigma(\bar \rho(y))}{D(\bar \rho(y))}\delta(y-y')
\nonumber \\
& &
+\epsilon T\frac{\bar J^2 A(\rho_0)}{D(\rho_0)^3}
\left[
\frac{y'(1-y)}{2}\theta(y-y')+(y \leftrightarrow y')\right].
\label{corr}
\end{eqnarray}
In order to present  a simpler demonstration, we 
calculate the intensity of the fluctuations of the
spatially averaged density
\begin{equation} 
\chi=\frac{1}{\epsilon} 
\bra \left( \int_0^1 dx \phi(x,t) \right)^2 \ket .
\end{equation}
Using (\ref{corr}), we obtain
\begin{equation}
\chi= T \int_0^1 dx \frac{\sigma(\bar \rho(x))}{D(\bar \rho(x))} 
+ T\frac{\bar J^2 A(\rho_0)}{24 D( \rho_0)^3 }.
\end{equation}
The first term on the right-hand side represents the contribution
of local equilibrium fluctuations, and the second term originates
from non-local fluctuations. For the special case 
$D=1$,  $\sigma=\rho(1-\rho)$, and $T=1$, the second term becomes 
$-(\rho_1-\rho_0)^2/12$,  which is consistent with the result 
obtained from the exact solution in Ref. \cite{DerridaLebowitzSpeer01}.

\section{Concluding remarks}\label{remark}


We have derived the large deviation functional 
for a simple model using fluctuating hydrodynamics. 
In particular, focusing on the deviation $N(\hat \rho())$ 
from the local equilibrium part, we obtain its lowest 
order expression with respect to the average current 
$\bar J$. This expression provides us with the physical 
interpretation that $N(\hat \rho() )$ corresponds to
the excess entropy absorption during the most probable 
process of generating the fluctuation $\hat \rho()$ at $t=0$ 
starting from $\bar \rho(x)$ at $t=-\infty$. 


The evolution equation describing the most probable process 
generating the fluctuation $\hat \rho()$ at $t=0$ is called 
the adjoint hydrodynamic equation. Within our approximation, 
we have obtained the adjoint hydrodynamic equation as 
(\ref{r-revol-hat}). In contrast to equilibrium cases,
the generating process is not given by the time reversal of 
a relaxation process  owing to the existence of the 
second term in  (\ref{r-revol-hat}). Rather, it may be noticed 
that (\ref{r-revol-hat}) is similar to the equation for driven 
diffusion systems. It is an interesting subject to elucidate
the physical picture for this asymmetry.


Our results can be generalized to those for several cases 
in a straightforward manner. For example, the  calculation 
of the large deviation functionals for driven diffusive 
systems may be a natural problem, which will be studied in 
a similar manner. Furthermore, the analysis of fluctuating 
hydrodynamics for a simple fluid is a physically important 
problem \cite{Eyink, Schmitz}. With regard to this problem, 
it has recently been found that the 
stationary distribution for non-equilibrium systems 
with multiple heat reservoirs is expressed in terms 
of the excess heat up to the order of the square of the average heat 
flux \cite{Komatsu}. The connection between the two 
results will be explored. 


Before concluding this paper, let us recall again that 
large deviation functionals in equilibrium cases are 
expressed by a thermodynamic function.  Thus,  
we are naturally led to consider a thermodynamic framework 
consistent with the expression of the large deviation  functional 
that we have calculated in this paper. For example, by using 
a method similar to that developed in Ref. \cite{HatanoSasa01}, 
we can derive an identity for the operations expressed by a 
parameter change. Applying the identity to the case in which
the boundary conditions are changed, one may find some insight 
related to an extended framework of thermodynamics. 

Furthermore, we wish to highlight the fact that the additivity 
of the system is an important property in a thermodynamic 
framework \cite{SST}. However, the additivity at a 
thermodynamic level is not directly related to 
the additivity of large deviation functionals
in non-equilibrium steady states  because of the 
existence  of non-local fluctuations. Therefore, one
may study the additivity 
principle that was discovered for exactly solvable models
\cite{DerridaLebowitzSpeer01,
DerridaLebowitzSpeer02b,
BodineauDerrida04} from the viewpoint of 
an extended framework of thermodynamics. 
The concrete expression given in (\ref{Nexp2}) will help
us to consider such a  problem.


\ack
The author thanks T. S. Komatsu  and H. Tasaki for related
discussions. This work was supported by a grant (No. 19540394) from 
the Ministry of Education, Science, Sports and Culture of Japan. 


\end{document}